\documentclass[sigconf]{acmart}
\AtBeginDocument{%
  }

\copyrightyear{2025}
\acmYear{2025}
\setcopyright{acmlicensed}\acmConference[UbiComp Companion '25]{Companion of the 2025 ACM International Joint Conference on Pervasive and Ubiquitous Computing}{October 12--16, 2025}{Espoo, Finland}
\acmBooktitle{Companion of the 2025 ACM International Joint Conference on Pervasive and Ubiquitous Computing (UbiComp Companion '25), October 12--16, 2025, Espoo, Finland}
\acmDOI{10.1145/3714394.3756218}
\acmISBN{979-8-4007-1477-1/2025/10}

\acmConference[UbiComp/ISWC 2025]{ACM International Joint Conference on Pervasive and Ubiquitous Computing}{October 12--16,
  2025}{Aalto university, Finland}

\usepackage{multirow}
\usepackage[acronym]{glossaries}
\makeglossaries
\newacronym{ap}{AP}{Access Point}
\newacronym{cir}{CIR}{Channel Impulse Response}
\newacronym{cnn}{CNN}{Convolutional Neural Network}
\newacronym{csi}{CSI}{Channel State Information}
\newacronym{dl}{DL}{Downlink}
\newacronym{iot}{IOT}{Internet of Things}
\newacronym{json}{JSON}{Java Script Object Notation}
\newacronym{ml}{ML}{Machine Learning}
\newacronym{mcs}{MCS}{Modulation and Coding Scheme}
\newacronym{mimo}{MIMO}{Multiple-Input  Multiple-Output}
\newacronym{nlos}{NLOS}{Non Line-of-Sight}
\newacronym{ota}{OTA}{Over-the-Air}
\newacronym{ofdm}{OFDM}{Orthogonal Frequency Division Multiplexing}
\newacronym{pars}{PARS}{Pattern Recofigurable Antenna Systems}
\newacronym{ppdu}{PPDU}{PHY Protocol Data Unit}
\newacronym{rf}{RF}{Radio Frequency}
\newacronym{sta}{STA}{Station Device}
\newacronym{ul}{UL}{Uplink}
\newacronym{xgboost}{XGBoost}{Extreme Gradient Boosting}

\begin{document}

\title{WiFi-Based People Counting Using Beam-Steerable Antennas: A Test-bed Study}


\author{Riccardo Bersan}
\email{riccardo.bersan@adant.com}
\affiliation{%
  \institution{Adant Technologies Inc.}
  \city{Padova}
  \country{Italy}
}
\author{Anay Ajit Deshpande}
\email{anay.deshpande@adant.com}
\orcid{}
\affiliation{%
  \institution{Adant Technologies Inc.}
  \city{Padova}
  \country{Italy}
}

\author{Sanaz Kianoush}
\email{sanaz.kianoush@cnr.it}
\affiliation{%
  \institution{IEIIT institute, Consiglio Nazionale delle Ricerche}
  \city{Milan}
  \country{Italy}
}

\author{Daniele Piazza}
\email{daniele.piazza@adant.com}
\affiliation{%
  \institution{Adant Technologies Inc.}
  \city{Padova}
  \country{Italy}
}

\author{Stefano Savazzi}
\email{stefano.savazzi@cnr.it}
\affiliation{%
  \institution{IEIIT institute, Consiglio Nazionale delle Ricerche}
  \city{Milan}
  \country{Italy}}

\renewcommand{\shortauthors}{Riccardo Bersan, Anay Ajit Deshpande, Sanaz Kianoush, Daniele Piazza, \&Stefano Savazzi}

\begin{abstract}
  \gls{rf} sensing is an emerging technology paradigm that repurposes existing wireless communication networks, such as WiFi, for imaging and computer vision applications, namely ambient and human sensing. Recent research has demonstrated that \gls{rf} holography techniques can surpass human vision capabilities in several tasks such as identifying and resolving individuals within dense crowds, interpreting gestures and emotions, and capturing images through walls. The paper explores the use of pattern-reconfigurable antenna devices with unmodified WiFi signals for indoor people discrimination, namely counting the number of people co-present in the space. We discuss signal modeling, integration of beam-steering technology, and adaptation for this purpose. Initial case studies and analyses within a test-house environment are also presented.
\end{abstract}




\begin{teaserfigure}
  \includegraphics[width=\textwidth]{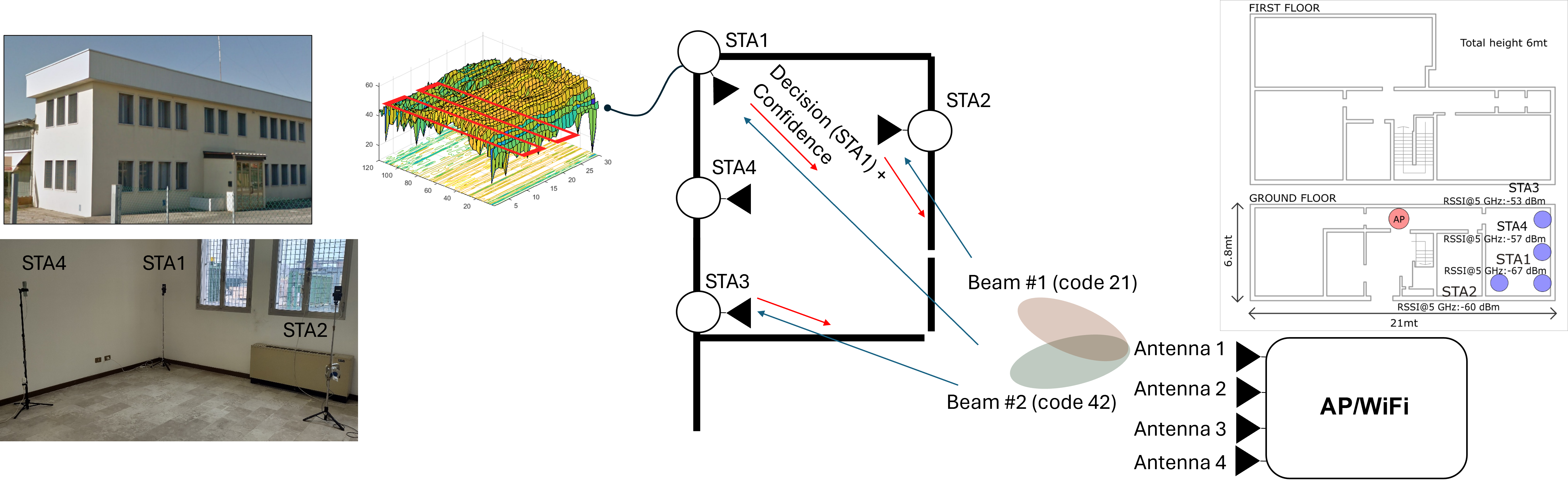}
  \caption{Wi-Fi data collection system for counting use case: ADANT testhouse environment}
  \Description{Wi-Fi data collection system for counting use case: ADANT testhouse environment}
  \label{fig:teaser}
\end{teaserfigure}


\maketitle

\section{Introduction}
\glsresetall
Ubiquitous perception through \gls{rf} signals is a pivotal opportunity for future technology \cite{youssef1}: it enables personalized services such as smart living, remote healthcare, automated logistics or interaction through free-space gestures \cite{mag1,Computer}. The ubiquity of Wi-Fi and cellular networks presents a promising platform for the development of innovative sensing tools. Future standards will also introduce dedicated sensing features which, for example, will allow routers to work as frequency modulated continuous wave radios targeting radar applications \cite{wifiradar}. Most of the current chip designs support ad-hoc firmware for \gls{csi} extraction with \gls{mimo} arrangements of the transmitter (TX) and receiver (RX) antennas and \gls{ofdm} subcarriers \cite{counting}. The \gls{csi} describes the phase shift and amplitude attenuation
of multiple propagation paths on each subcarrier. The latest IEEE 802.11be standard (Wi-Fi 7) offers a wider subcarrier bandwidth of $160$MHz (up to $320$MHz), providing at least $120$ usable pilot subcarriers for \gls{csi} or \gls{cir} estimation. Additionally, Wi-Fi signals have been recently exploited to track daily human movements and behaviors\cite{wificounting}, while Wi-Fi signal variations have been shown to differ between different people \cite{avola} and can consequently be used for their re-identification. 
\begin{figure*}
  \centering
  \includegraphics[scale=0.40]{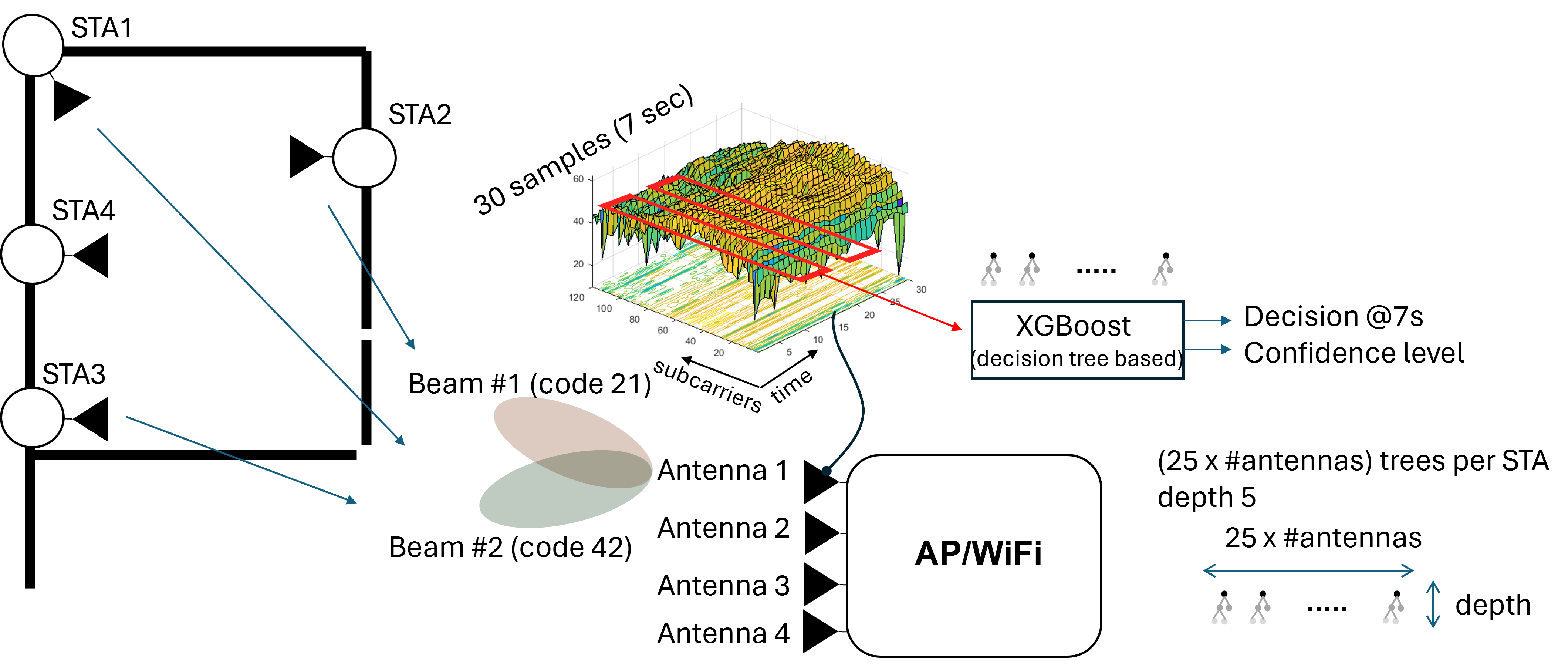}
  \caption{Uplink (UL) WiFi CSI data processing and infrastructure example.}
  \Description{Uplink (UL) WiFi CSI data processing and infrastructure example.}
  \label{fig:ul}
\end{figure*}

In recent times, \gls{pars} technologies have been proposed as a novel opportunity for human-scale passive sensing \cite{beamsteering}. Beam management techniques specifically beam steering and beam switching allows to dynamically channelize (or steer) the individual antenna radiation patterns to intended areas. As depicted in Fig.~\ref{fig:teaser}, each AP antenna system independently provides 2 steerable beams, indicated by beampatterns $21$ and $42$, respectively. The beam-steering system is deployed on the WiFi \gls{ap} while the beam management is provided through software control exploiting real-time selection and channelization.
Conventional beamforming and  smart-antenna systems need precise phase alignment and covariance matrix estimation to determine the optimal beamforming design. But in \gls{pars}, the
beam patterns are dynamically altered on each antenna separately. and do not require precise phase alignment simplifying product
integration which opens unprecedented opportunities for scaling up \gls{rf} sensing systems as well as for integration of communication and sensing services.

Taking into account the advantages of \gls{pars}, the paper proposes a Wifi sensing based people counting algorithm that provides a comparative evaluation of the performance of an \gls{rf} sensing system which supports beamsteering technology. Particularly, we propose a machine learning based beam-space processing architecture designed to split
the learning functions between the deployed STA devices and the AP.
The STA devices analyze the \gls{csi} observed over different antenna beampatterns and utilize an \gls{xgboost} decision tree model \cite{xgboost} for supervision.
Concurrently, a \gls{cnn} is trained on the
\gls{ap} to generate a final decision, based on the outputs of the local \gls{xgboost} trees. Validation tests were
conducted within a smart home environment: specifically, a large 340
sqm test house (see Fig. \ref{fig:teaser} on the left). The tests
aimed to achieve two primary objectives: first, to develop an enhanced
residential WiFi AP gateway with beamsteering capabilities; and second,
to provide motion detection and people counting services via the same
residential WiFi test network, supporting various application verticals
such as intrusion detection and building automation \cite{ubiloc}.

\section{Wi-Fi data collection system}
The paper focuses on the uplink communication of a WiFi communication system consisting of $N$ stations, or \glspl{sta} (namely, STA $i=1$, $i=2$, $i=3$, $i=4$) equipped
with $M_{T}$ antennas and communicating with one WiFi 7 compliant
access point (AP) gateway equipped with $M_{R}=4$ antennas with spacing
$D$. We resort to a practical case where the AP gateway is equipped with a smart antenna system and supports beam-steering functions, while the WiFi $N=4$ \glspl{sta} are equipped with a single antenna, $M_{T}=1$. The goal is to reconstruct an \emph{infrastructural snapshot}~\cite{moral} of the environment, namely, to obtain an accurate prediction of the number $\mathrm{X}$ of human subjects (i.e., the targets) moving in the monitored space as well as to capture the body motion patterns on a given time instant $t$.

The \gls{ap} is a commercially available device currently used by a large Internet Service Provider for its subscribers. It is based on the Broadcom BCM4912, a quad core 64-bit ARM processor, and a combination of three Broadcom BCM6715 WiFi6E radios, which concurrently cover all the 2.4GHz, 5GHz and 6GHz bands with 4x4 \gls{mimo} antenna systems and channel bandwidth up to 160MHz. 
This \gls{ap} is equipped with smart antenna technologies in the 5GHz and 6GHz bands, by incorporating \gls{rf} switches between the \gls{rf} chains and the antenna elements. The design allows for selecting and combining the radiation patterns for different antenna elements placed in the devices, achieving isotropic coverage at each \gls{rf} chain and effectively covering the holes in radiation patterns that are typically left by the passive antenna design.

The proposed application case study is meant to design a system that provides augmented sensing functions for real-time tracking of the
number of subjects co-present and moving in different areas of the
testhouse environment. As depicted in Fig. \ref{fig:teaser}, the WiFi \gls{sta} devices and the WiFi \gls{ap} gateway
are in \gls{nlos} and located in two different rooms.
Regarding data collection and processing, the \gls{csi} collection duration for training corresponds to $30$ minutes for each test, with sampling frequency of $250$ ms, corresponding to $4$ CSI samples/sec for each \gls{sta} (which corresponds to around ~30K samples for each test). 

Body motions are detected by inspection and real-time analysis of the \gls{csi} response $\mathbf{H}_{t}(i,k)$ observed by the \gls{sta} $i$ and corresponding to \gls{ap} antenna beampattern mode $k$, at discrete time instants $t=1,2,...$. Each time instant corresponds to a WiFi \gls{ppdu} frame. The \gls{csi} matrix $\mathbf{H}_{t}(i,k)\in\mathbb{C}^{M_{R}\times S}$ observed by the \gls{sta} device at time $t$ has thus dimension $M_{R}\times S$ with $S=120$ being the number of usable pilot subcarriers.

As discussed previously, the goal of the tests is to determine the room occupancy, namely to estimate the density of people occupying an assigned room.
During training and calibration stages, we recorded the \gls{csi} corresponding to a single individual moving in the selected area (Figure \ref{fig:teaser}) and gradually increase the number of individuals co-present in the same area up to a maximum of $\mathrm{X=8}$ subjects. 
As the number of subjects increases, the body-induced
blockage affects the \gls{csi} of more subcarriers, resulting in distinct \gls{rf} fingerprints.
During testing, we record people walking, standing, or moving in groups in the same area.

The \gls{ap} serves as edge node and is responsible not only for collecting the \gls{csi} but also for its processing. Besides the \gls{ap}, the proposed architecture allows the \gls{sta} nodes to deploy and execute simplified models, such as \gls{xgboost}, for local computations, as further elaborated in the subsequent discussion. In the proposed settings,
the evaluation of the proposed beam-space processing architectures 
are implemented on a remote PC: the \gls{csi} features are moved/transmitted
by using \gls{json} serialization and the
MQTT publisher/subscriber transport service \citep{mqtt}.

\begin{figure*}
  \centering
  \includegraphics[scale=0.40]{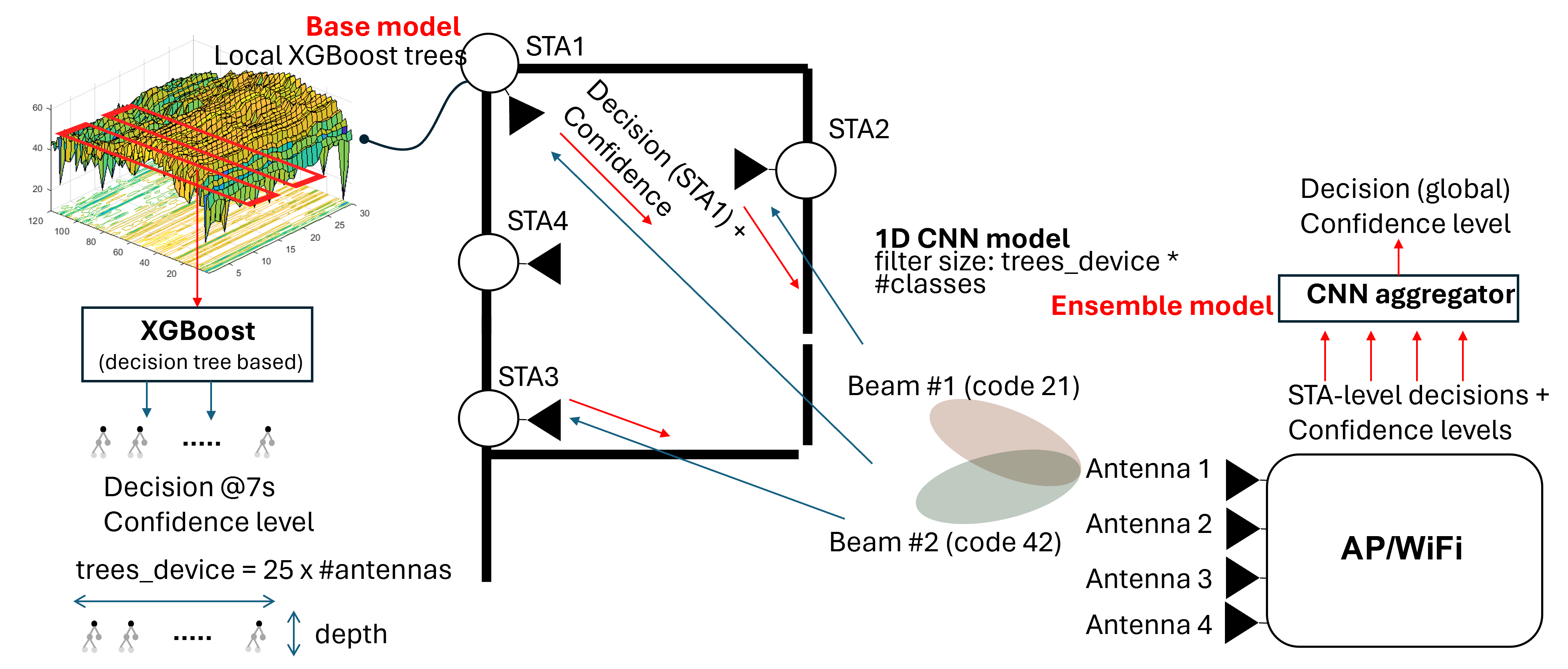}
  \caption{\gls{dl} WiFi CSI data processing and split learning framework example.}
  \Description{\gls{dl} WiFi CSI data processing and split learning framework example.}
  \label{fig:dl}
\end{figure*}
\begin{table*}[tp]
\protect\caption{\label{processing_scenarios}CSI processing scenarios}

\begin{centering}
\begin{tabular}{l|l|}
\hline 
\textbf{Scenario } & \textbf{Description} \tabularnewline
\hline 
\multirow{2}{*}{STA $i$, beam $k=21$} & Counting and subject discrimination are achieved by processing the
\gls{csi} matrix $\mathbf{H}_{t}(i,k)$ observed by \tabularnewline
 & STA $i=1,2,3,4$ and beam code $k=21$\tabularnewline
\hline 
\multirow{2}{*}{STA $i$, beam $k=42$} & Counting and subject discrimination are achieved by processing the
\gls{csi} matrix $\mathbf{H}_{t}(i,k)$ observed by \tabularnewline
 & STA $i=1,2,3,4$ and beam code $k=42$\tabularnewline
\hline 
\multirow{3}{*}{STA $i$, all beams} & Counting and subject discrimination are achieved by processing the
\gls{csi} matrix \tabularnewline
 & $\mathbf{H}_{t}(i)=[\mathbf{H}_{t}(i,k=21),\mathbf{H}_{t}(i,k=42)]$
obtained by AP \tabularnewline
 & RX antenna switching between beam codes $k=21$ and $k=42$ by time division\tabularnewline
\hline 
\multirow{2}{*}{UL-F} & Counting and subject discrimination are achieved by fusing the \gls{csi}
matrices \tabularnewline
 & $\mathbf{H}_{t}=[\mathbf{H}_{t}(i=1),...,\mathbf{H}_{t}(i=4)]$ obtained
by all the deployed STA on the AP\tabularnewline
\hline 
\multirow{3}{*}{DL-SLs} & A decision $\widehat{\mathrm{X}}(i,k)$ about the number $N$ of subjects co-present
in the room is made locally by each \tabularnewline
 & STA $i$ and observed beam $k$. The results are collected by the AP into a matrix $\widehat{\mathbf{X}}=\left\{ \widehat{\mathrm{X}}(i,k=21),\widehat{\mathrm{X}}(i,k=42)\right\} _{i=1}^{4}$ \tabularnewline
 & and fed to a second processing stage (aggregation stage) and a final decision.\tabularnewline
\hline 
\end{tabular}
\par\end{centering}
\medskip{}
\end{table*}

\section{CSI processing infrastructure}

Table \ref{processing_scenarios} summarizes the \gls{csi} processing scenarios
considered in the testbed study. For all cases the \gls{csi} matrix $\mathbf{H}_{t}(i,k)$
is collected over time for each STA $i,$ and beam steering profile
(2 beampatterns per antenna, namely $k=21$ and $k=42$). Information about the time-varying \gls{mcs} is also available for each \gls{ppdu} frame. The approach we followed is to detect
and classify the variations of the estimated \gls{csi} amplitude samples
over a sliding window of duration $\bigtriangleup T=7$ consecutive
WiFi \gls{ppdu} frames. This is calculated as follows:

\begin{equation}
\triangle\mathbf{H}_{t}(i,k)=\sqrt{\frac{1}{\Delta T}\sum_{\tau=t-\Delta T+1}^{t}\left[\left|\mathbf{H}_{\tau}(i,k)\right|-\mathbf{\boldsymbol{\mu}}_{\tau}(i,k)\right]^{2}},\label{eq:features}
\end{equation}
with $\mathbf{\boldsymbol{\mu}}_{t}(i,k)=\frac{1}{\Delta T}\sum_{\tau=t-\Delta T+1}^{t}\left|\mathbf{H}_{t}(i,k)\right|$.
In addition to the scenarios described in Table \ref{processing_scenarios}, we highlight two beam-space
data processing architectures, described as follows. 
\begin{itemize}
\item \textbf{UL-F: Uplink CSI data fusion on the AP}. The first case,
shown in Fig. \ref{fig:ul}, involves data collection at the Wi-Fi \gls{ap} (over
uplink) and the subsequent training of a single
\gls{ml} model. An XGBoost model is trained on the AP considering all the
\gls{csi} variations $\triangle\mathbf{H}_{t}(i,k)$ as inputs, $\forall i,k$.
\item \textbf{DL-SL: Downlink split learning}. The second case, depicted
in Fig. \ref{fig:dl}, proposes a split learning architecture where a portion
of the \gls{ml} model is trained locally on each Wi-Fi \gls{sta} device. The results
of this processing are collected by the Wi-Fi \gls{ap} and then fed to a
second aggregation stage. Thus, a second algorithm/model is trained independently.
The split learning architecture is attractive due to its scalability
potential and support to massive \gls{rf} data processing, since raw \gls{csi} are kept on the STA device and never shared \cite{sensors_sanaz}. The chosen local
model trained on the \glspl{sta} is the \gls{xgboost}, while the model used for
aggregating the partial decisions of each \gls{sta} is a \gls{cnn}.
\end{itemize}
Table \ref{parameters} summarizes the main parameters of the local model (running on the \gls{sta}) and global/aggregation
model (on the \gls{ap}) used for UL-F and DL-SL architectures.

\begin{table}[tp]
\protect\caption{\label{parameters} UL-F and DL-SL models: local and aggregator model}
\vspace{0cm}

\begin{centering}
\begin{tabular}{|c|c|c|}
\cline{2-3} 
\multicolumn{1}{c|}{} & \multicolumn{1}{c|}{\textbf{UL-F}} & \multicolumn{1}{c|}{\textbf{DL-SL}}\tabularnewline
\hline 
\textbf{Inputs} & $\triangle\mathbf{H}_{t}(i,k),$$\forall i,k$  & $\triangle\mathbf{H}_{t}(i,k),$$\forall i,k$\tabularnewline
\hline 
\multirow{3}{*}{\textbf{Local Model}} & \multirow{3}{*}{n.a.} & XGBooost\tabularnewline
 &  & trees per STA: $25\times M_{R}$\tabularnewline
 &  & depth: $5$\tabularnewline
\hline 
\multirow{3}{*}{\textbf{Aggregator (AP)}} & XGBooost & 1D CNN layer\tabularnewline
 & trees: $25\times4\times M_{R}$  & filter size: $25\times8$\tabularnewline
 & depth: $5$ & \tabularnewline
\hline 
\end{tabular}
\par\end{centering}
\medskip{}
 \vspace{-0.6cm}
\end{table}

\section{Test-house environment description}
In this section, we introduce the test-house infrastructure of ADANT and outline the tests for the measurement campaign. Here, we introduce the test types, set up, and configurations of the \gls{csi} processing infrastructure.
WiFi \gls{csi} data are collected in the testhouse environment for evaluating the algorithm and optimized recognition beams. The goal is to recognize specific movements and locations and improve machine learning algorithms for data classification.
The testhouse environment (see Fig. \ref{fig:teaser}) is located in Selvazzano Dentro - Padova (IT) and consists of a two-story structure with an area of approximately 340 sqm. In particular, the following features are highlighted:
pristine Wi-Fi spectrum free of interference, 
state-of-the-art equipment for fully automated \gls{ota} testing for Wi-Fi and \gls{iot}, compliant with various Wi-Fi standards, including Wi-Fi 6, 6E, and 7. The facility includes eleven rooms and a corridor on each floor that can be used to emulate different types and ranges of installations, ensuring maximum reliability and reproducibility of the results.

\section{Results}
\begin{figure}
  \centering
  \includegraphics[scale=0.24]{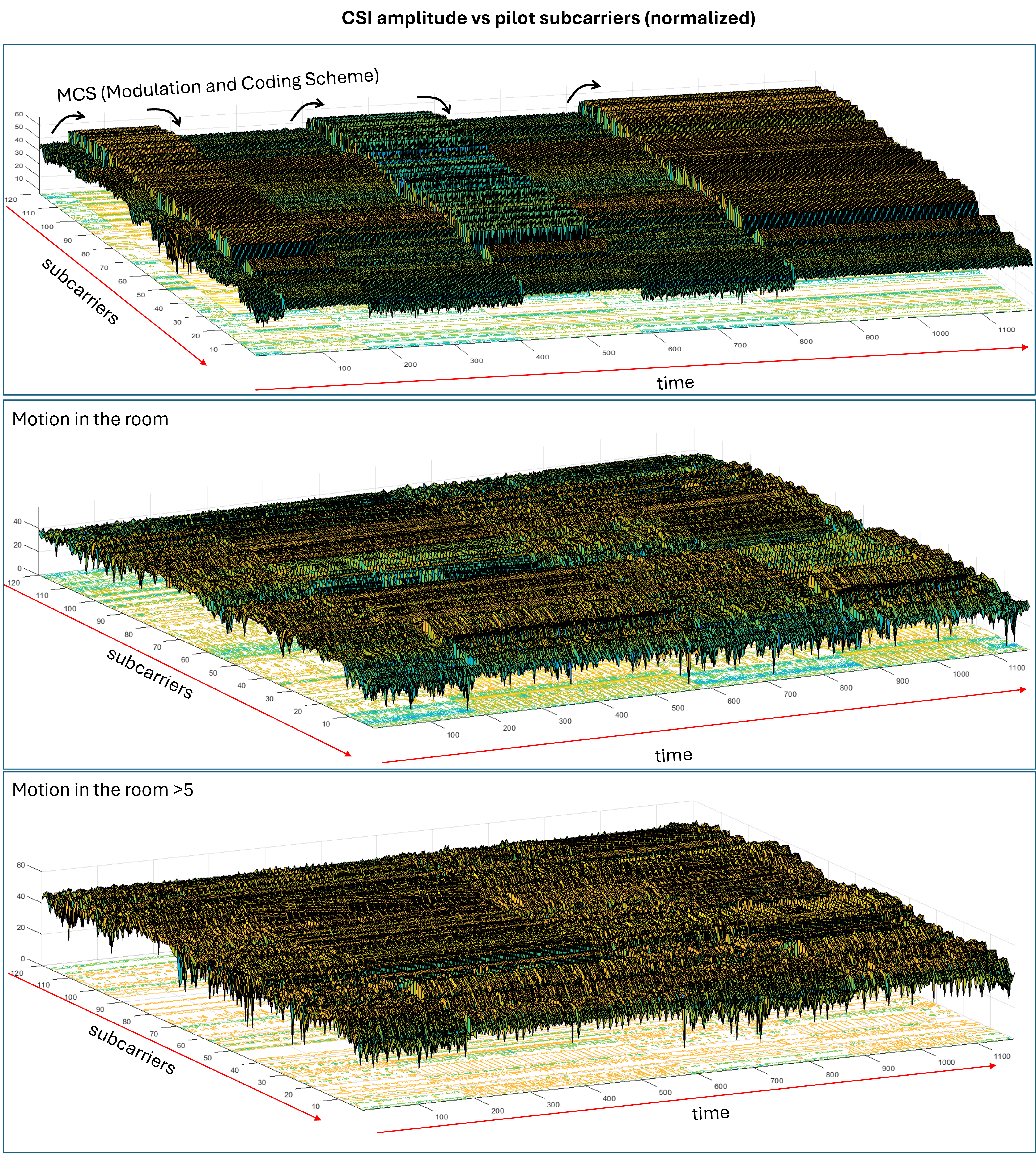}
  \caption{CSI example data: normalized amplitude $\left|\mathbf{H}_{t}(i,k)\right|$ vs the $S=120$ pilot subcarriers for empty environment and different motions example. \gls{mcs} change over consecutive \gls{ppdu} frames.}
\Description{CSI example data: normalized amplitude vs pilot subcarriers for empty environment and different motions example. Modulation and Coding Scheme (MCS) change over consecutive PPDU frames.}
  \Description{CSI example data: normalized amplitude vs pilot subcarriers for empty environment and different motions example. Modulation and Coding Scheme (MCS) change over consecutive PPDU frames.}
  \label{fig:csi_samples}
\end{figure}

Figure \ref{fig:csi_samples} shows some examples of the observed normalized
\gls{csi} amplitudes $\left|\mathbf{H}_{t}(i,k)\right|$ in log scale across
the $S=120$ subcarriers and for a consecutive number of \gls{ppdu} frames ($t$).
Three cases presented are: 
\begin{enumerate}
    \item empty testhouse environment with no subjects moving in the considered room;
    \item $\mathrm{X}=1$ subject in motion;
    \item $\mathrm{X}\geq5$ co-present subjects.
\end{enumerate}
The sensitivity of the \gls{csi} to the number of targets perturbing the link between \gls{sta} $i=1$
and the \gls{ap} node is noticeable. The abrupt changes observed in the
\gls{csi} samples within the figure are attributed to the AP node re-selection
of the \gls{mcs}. While this reselection
is beyond the control of the \gls{csi} processing tool, its effects are
minimized by calculating the variations in estimated \gls{csi} amplitude
samples across consecutive \glspl{ppdu} as $\triangle\mathbf{H}_{t}(i,k)$.

\begin{figure}
  \centering
  \includegraphics[width=\linewidth]{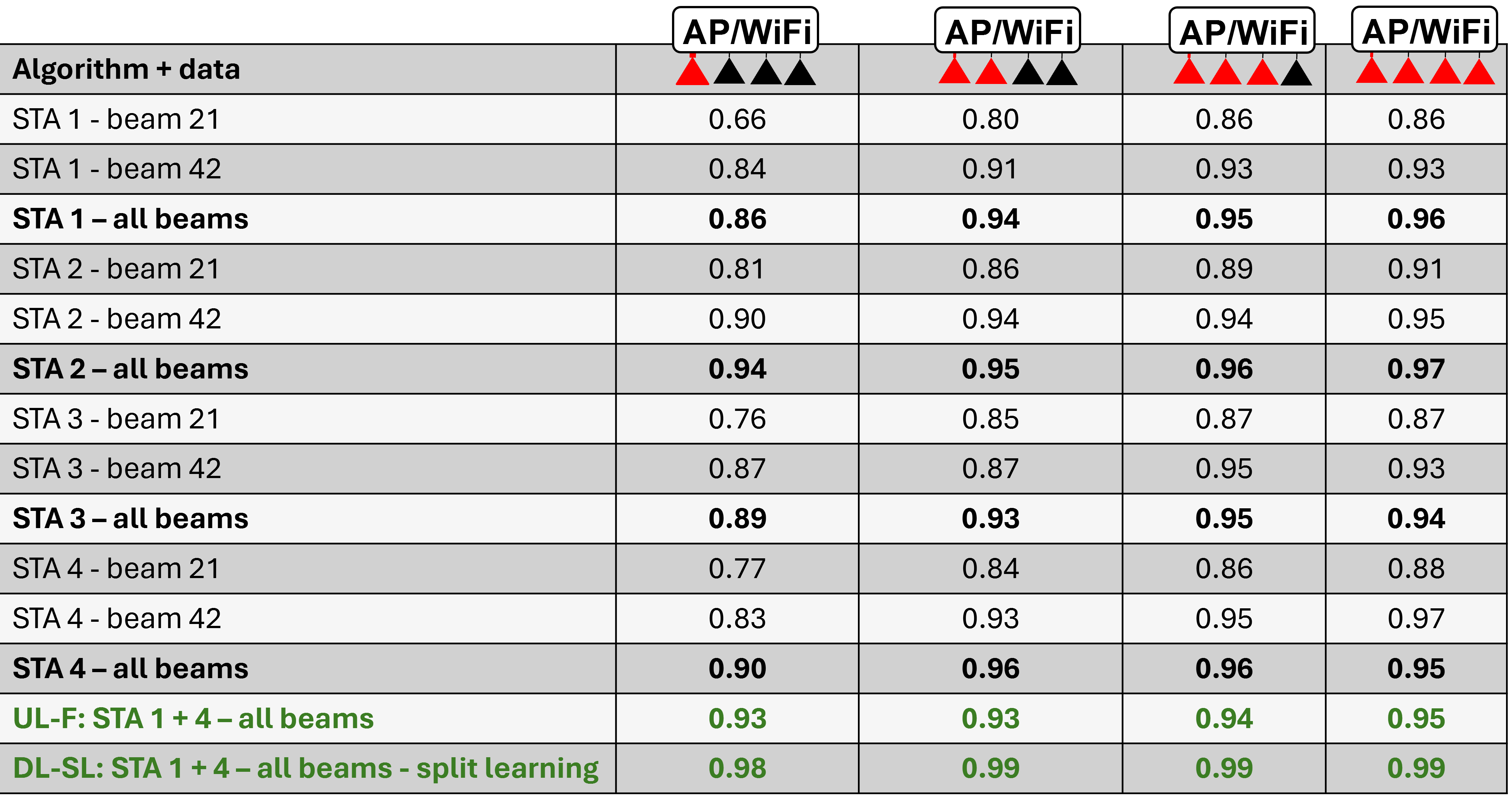}
  \caption{Counting accuracy measured on all WiFi station devices (STA 1-4), for all beampatterns (beam 21 and 42) and varying number of RX antennas at the AP node. Results are highlighted for $\textit{X=8}$ subjects co-present. Performance of UL-F and DL-SL processing pipelines are highlighted in green.}
  \Description{Counting accuracy measured on all WiFi station devices (STA 1-4), for all beampatterns (beam 21 and 42) and varying number of RX antennas at the AP node. Results are highlighted for $\textrm{X=8}$ subjects co-present. Performance of UL-F and DL-SL processing pipelines are highlighted in green.}
  \label{fig:acc_RX}
\end{figure}

\begin{figure}
  \centering
  \includegraphics[width=\linewidth]{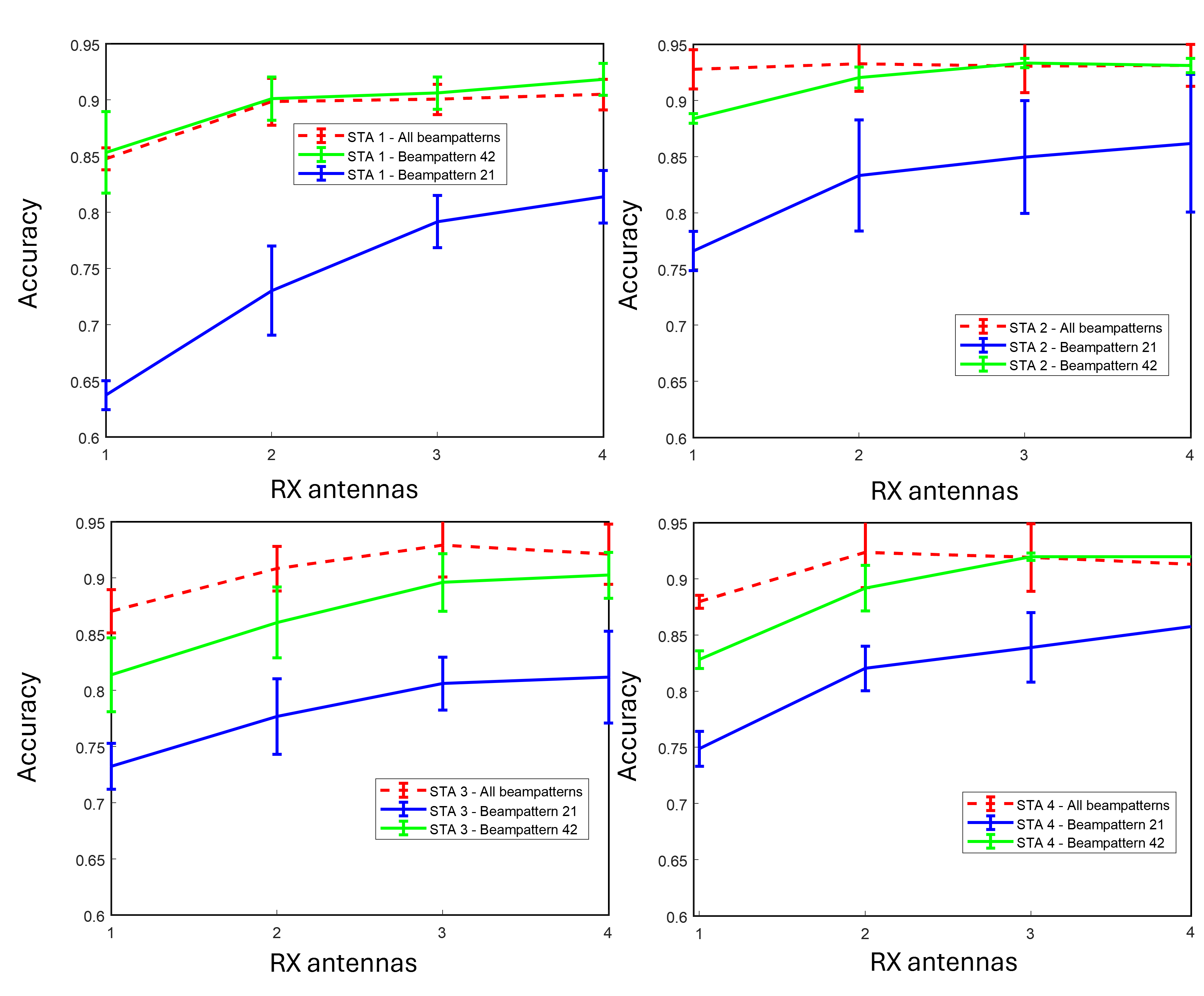}
  \caption{Average accuracy vs the number of RX antennas at the AP for processing taken place on STA 1-4 separately. Green lines correspond to beampattern code 42, blue lines correspond to beampattern code 21. Dashed lines apply the joint use of both patterns. Error bars are superimposed for each case according to a three-fold cross-validation.}
  \Description{Average accuracy vs the number of RX antennas at the AP for processing taken place on STA 1-4 separately. Green lines correspond to beampattern code 42, blue lines correspond to beampattern code 21. Dashed lines apply the joint use of both patterns.}
  \label{fig:avg_acc}
\end{figure}

\begin{figure*}
  \centering
  \includegraphics[scale=0.40]{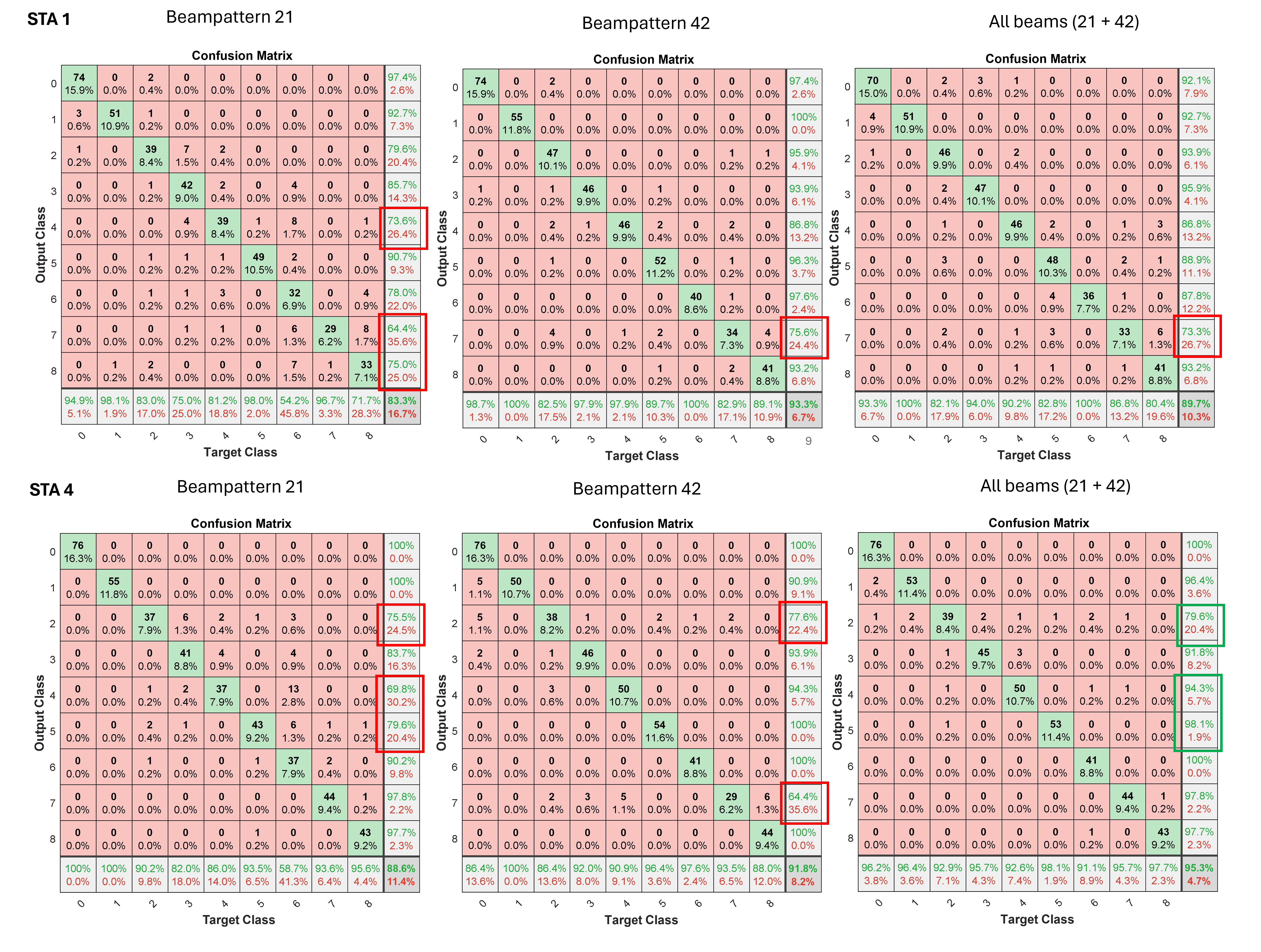}
  \caption{Confusion matrices for the counting problem ($\mathrm{X=8}$ subjects): (a) STA $i=1$, beampattern code $k=21$; (b) STA $i=1$, beampattern $k=42$; (c) STA $i=1$, beampatterns codes $k=21$ and $k=42$; (d) STA $i=4$, beampattern code $k=21$; (e) STA $i=4$, beampattern $k=42$; (f) STA $i=4$, beampatterns codes $k=21$ and $k=42$.}
  \Description{Confusion matrices for the counting problem: (a) STA 1, beampattern code 21; (b) STA 1, beampattern 42; (c) STA 1, beampatterns codes 21 and 42; (d) STA 4, beampattern code 21; (e) STA 4, beampattern 42; (f) STA 4, beampatterns codes 21 and 42.}
  \label{fig:confusion}
\end{figure*}

The table of Fig. \ref{fig:acc_RX} summarize the counting accuracy
obtained by processing the \gls{csi} variations $\triangle\mathbf{H}_{t}(i,k)$ on individual WiFi \glspl{sta} (\gls{sta} $i=1,2,3,4$) and using one RX antenna at the AP node. For each \gls{sta}, we gather the \gls{csi} information on the two beampatterns (code $k=21$ and $k=42$) separately and jointly, for computing the counting accuracy , i.e., beampatterns are switching between beam code $k=21$ and $k=42$ according to a time division policy. Accuracy is also computed for varying number $(\textrm{X}$) of subjects co-present, where $\textrm{X}=1,...,8$. Finally, the counting performance of UL-F and DL-SL processing pipelines are highlighted in green. From the figure, there is a clear degradation in average performance for beampattern $21$ compared to beampattern $42$. However, the combined use of both beampatterns is effective in the majority of cases and contributes to increasing the average accuracy by approximately 10\%, considering, for example, the counting of $\textrm{X}=8$ moving
targets. The DL-SL architecture appears to perform better than the
UL-F approach, which involves \gls{csi} data fusion at the AP. As shown in Fig. \ref{fig:acc_RX}, increasing the number of RX antennas, i.e., from $M_{R}=1$ to $M_{R}=4$ improves the accuracy as expected while accuracy degradation on beampattern $21$ (compare to beam code $42$) is still observed. Figs. \ref{fig:avg_acc} and \ref{fig:confusion} analyze in detail the increase in performance that can be achieved from the combined use of the two available beampatterns for each \gls{ap} antenna. Fig. \ref{fig:avg_acc} shows the accuracy obtained by combining all beams, compared to that achievable using separate beams (green and blue lines). The corresponding confusion matrices are analyzed in detail in Fig. \ref{fig:confusion} for \gls{sta} $i=1$ and \gls{sta} $i=4$, respectively. The benefits derived from the joint use of both beampatterns are highlighted (in green) and are particularly noticeable when the number of targets to be discriminated is high ($\mathrm{X}>4$).

\begin{figure}
  \centering
  \includegraphics[width=\linewidth]{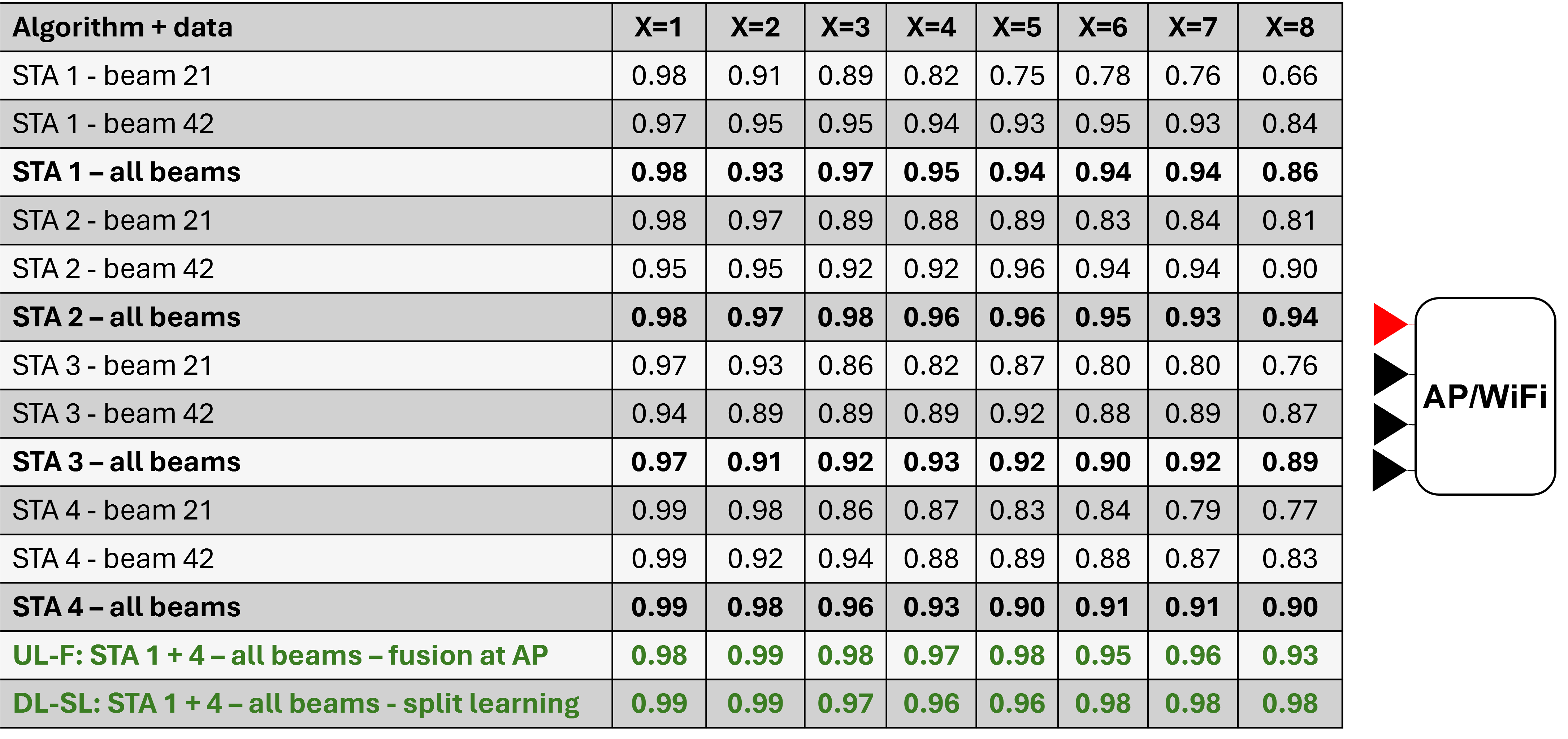}
  \caption{Counting accuracy measured on all WiFi station devices (STA 1-4),
for all beampatterns (beam codes $k=21$ and $k=42$) and varying number $\textrm{X}$
of subjects co-present. The performance of UL-F and DL-SL processing
pipelines are also highlighted in green. In these examples the \gls{csi}
is collected from one RX antenna at the AP.}
  \Description{Counting accuracy measured on all WiFi station devices (STA 1-4),
for all beampatterns (beam codes $k=21$ and $k=42$) and varying number $\textrm{X}$
of subjects co-present. The performance of UL-F and DL-SL processing
pipelines are also highlighted in green. In these examples the CSI
is collected from one RX antenna at the AP.}
  \label{fig:acc_x}
\end{figure}

\section{Concluding remarks}

The paper introduced the use of the beam-steering antenna technology
for people counting using ambient
WiFi signals. \gls{pars}s can channelize the radiation
energy to improve the coverage over selected areas and are thus helpful
for environment-aware joint sensing and communication applications.
Two models for beam-space processing have been proposed and applied. Validation tests have been carried out inside a test-house environment with the purpose of verifying the robustness of the counting system inside a representative residential WiFi network.
\bibliographystyle{unsrt}
\bibliography{sample-base}

\end{document}